\documentclass[10pt,aps,twocolumn,prd,superscriptaddress,showpacs,nofootinbib,noshowkeys,floatfix,preprintnumbers]{revtex4}
\usepackage{times}

\usepackage[utf8]{inputenc}

\usepackage{graphics,graphicx}
\usepackage{amsmath, amssymb}
\usepackage{multirow}
\usepackage{longtable}
\usepackage{color}

\newcommand{\dd}{\mathrm{d}}

\usepackage[colorlinks=true,linktocpage=true,linkcolor=blue,citecolor=blue,hyperfootnotes=true]{hyperref}
\usepackage[normalem]{ulem}  
\usepackage{subfigure}

\begin{document}
	
	
	\title{Matching the Hagedorn mass spectrum with lattice QCD results}
	
	\author{Pok Man Lo}
	\affiliation{Institute of Theoretical Physics, University of Wrocław, PL-50204 Wrocław, Poland}
	\author{Michał Marczenko}
	\affiliation{Institute of Theoretical Physics, University of Wrocław, PL-50204 Wrocław, Poland}
	\author{Krzysztof Redlich}
	\affiliation{Institute of Theoretical Physics, University of Wrocław, PL-50204 Wrocław, Poland}
	\affiliation{Extreme Matter Institute EMMI, GSI, Planckstrasse 1, D-64291 Darmstadt, Germany}
	\affiliation{Department of Physics, Duke University, Durham, NC 27708, USA}
	\author{Chihiro Sasaki}
	\affiliation{Institute of Theoretical Physics, University of Wrocław, PL-50204 Wrocław, Poland}
	\affiliation{Frankfurt Institute for Advanced Studies, D-60438 Frankfurt am Main, Germany}
	
	\date{\today}

	\begin{abstract}
		Based on recent lattice QCD (LQCD) results obtained at finite temperature, we discuss modeling of the hadronic phase of QCD in the framework of hadron resonance gas (HRG) with discrete and continuous mass spectra. We focus on fluctuations of conserved charges, and show how a common limiting temperature can be used to constrain the Hagedorn exponential mass spectrum in different sectors of quantum number, through a matching of HRG and LQCD. For strange baryons, the extracted spectra are found to be consistent with all known and expected states listed by the Particle Data Group (PDG). The strange-mesonic sector, however, requires additional states in the intermediate mass range beyond that embodied in the database.
	\end{abstract}
	
	
	\maketitle

	\section{Introduction}
	\label{intro}

	The thermodynamics of the confined phase of QCD is commonly modeled with the hadron resonance gas (HRG)~\cite{BraunMunzinger:2003zd,tawfik1,tawfik2,reseqsf,andronic,kapusta,kapusta1,gorenstein}. The equation of state for strongly interacting matter at finite temperature is well described by this model, formulated with a discrete mass spectrum of the experimentally confirmed particles and resonances. This finding was verified by recent results of lattice QCD (LQCD)~\cite{Bazavov:2012jq, Borsanyi:2013bia,missing,Borsanyi:2011sw}. However, LQCD also reveals that, when considering fluctuations and correlations of conserved charges, there are clear limitations in the HRG description~\cite{missing}. This is particularly evident in the strange sector, where the second-order correlations with the net-baryon number $\chi_{\rm BS}$ or strangeness fluctuations $\chi_{\rm SS}$ are larger in LQCD than those in the HRG model~\cite{missing,Bazavov:2012jq}. Such deviations were attributed to the missing resonances in the Particle Data Group (PDG) database~\cite{missing}.
	
	Different extensions of the HRG model have been proposed to quantify the LQCD equation of state. They account for a possible repulsive interaction among constituents and/or a continuously growing exponential mass spectrum~\cite{andronic,gorenstein,Majumder:2010ik,kapusta}. The latter was first introduced by Hagedorn~\cite{Hagedorn:1965st} within the statistical bootstrap model (SBM)~\cite{Hagedorn:1971mc,Frautschi:1971ij,raf1}, and was then studied in dual string and bag models~\cite{Huang:1970iq,Cudell:1992bi,Johnson:1975sg}. For large masses, the Hagedorn spectrum $\rho(m)$ is parametrized as \mbox{$\rho(m)\simeq m^a e^{m/T_H}$}, where $T_H$ is the Hagedorn limiting temperature and $a$ is a model parameter.
	
	The main objective of this paper is to analyze LQCD data on fluctuations and correlations of conserved charges within the HRG model. In particular, we examine whether the missing resonances contained in the asymptotic Hagedorn mass spectrum are sufficient to quantify LQCD results. We focus on the susceptibilities $\chi_{\rm BS}$ and $\chi_{\rm SS}$, where LQCD indicates the largest deviations from HRG, in spite of their agreement on the equation of state in the hadronic phase.
	
	To calculate fluctuations of conserved charges within HRG, one needs to identify the hadron mass spectrum for different quantum numbers. For a continuous mass spectrum $\rho(m)$, this issue was addressed in Refs.~\cite{Broniowski:2000bj} and~\cite{Broniowski:2004yh}, where the parameters of $\rho(m)$ in different hadronic sectors were extracted by fitting the spectra to the established hadronic states in the PDG database~\cite{pdg}. It was shown in Ref.~\cite{Broniowski:2004yh} that the Hagedorn temperatures for mesons $T_H^M$ and baryons $T_H^B$ are different, with $T_H^M>T_H^B$. The $T_H^B\simeq 140 \, {\rm MeV}$ found in~\cite{Broniowski:2000bj} is clearly below the LQCD crossover temperature $T_c = 155(1)(8) \, {\rm MeV}$ from hadronic to quark-gluon plasma phase~\cite{tcb,tcw,tc}. This, however, is inconsistent with LQCD, as it implies a large fluctuation of the net-baryon number deep in the hadronic phase, which is not observed in lattice simulations. 
	
	In this study we have reanalyzed the Hagedorn mass spectrum in different sectors of quantum number, in the context of the PDG data, and have shown that there is a common Hagedorn temperature for mesons and baryons in different strange sectors. We have applied our newly calculated $\rho(m)$ in the HRG model to explore different thermodynamics observables; in particular, fluctuations of conserved charges. The results are compared with LQCD for the strangeness, net-baryon number fluctuations, and for baryon-strangeness correlations. We show that HRG, adopting a continuous mass spectrum with its parameters fitted to the PDG data, can partially account for the missing resonances needed to quantify LQCD results.
	
	To fully identify the missing resonance states, we motivate a matching of LQCD and HRG to extract a continuous mass spectrum $\rho(m)$. In the strange-baryonic sector, this $\rho(m)$ is shown to be consistent with all known and expected states listed by the PDG. However, the mass spectrum for strange mesons require some additional resonances in the intermediate mass range beyond those listed in the PDG compilation.
	
	The paper is organized as follows: In Sec.~\ref{HRG}, we introduce the HRG thermodynamics with a discrete mass spectrum. In Sec.~\ref{HRG_LQCD}, we discuss HRG model comparisons with LQCD. In Sec.~\ref{Hagedorn_LQCD}, we extract the continuous $\rho(m)$ in different sectors of quantum number and discuss fluctuations of conserved charges in conjunction with LQCD findings. Finally, Sec.~\ref{conclusions} is devoted to summary and conclusions.
	
	\section{Equation of state of hadronic matter}
	\label{HRG}
	
	To formulate a phenomenological model of hadronic matter at finite temperature and density, one needs to identify the relevant degrees of freedom and their interactions. In the confined phase of QCD the medium is composed of hadrons and resonances.
	
	The HRG model, in its simplest form, treats the medium constituents as point-like and independent~\cite{BraunMunzinger:2003zd}. Thus, in such a model setup, the interactions of hadrons and the resulting widths of resonances are neglected. Hence, the composition of the medium and its properties emerge through a discrete  mass spectrum,
	\begin{eqnarray}
	\label{DEF:rho_pdg}\label{eq2}
	\rho^{\rm{HRG}}(m) = \sum_{i} d_i \delta\left(m - m_i\right) \textrm{,}
	\end{eqnarray}
	where $d_i = (2J_i+1)$ is the spin degeneracy factor of a particle $i$ with mass $m_i$ and the sum is taken over all stable particles and resonances.
	
	The mass spectrum in Eq.~(\ref{eq2}) can be identified experimentally or can be calculated within LQCD. In both cases our knowledge is far from complete. LQCD can determine the masses of hadronic ground states and low-lying excited states with fairly high precision~\cite{spectrum}. However, the higher excited states are still not well controlled in lattice calculations.
	
	\begin{figure*}[htp!]
		\centering\subfigure[]{\includegraphics[width=1\columnwidth]{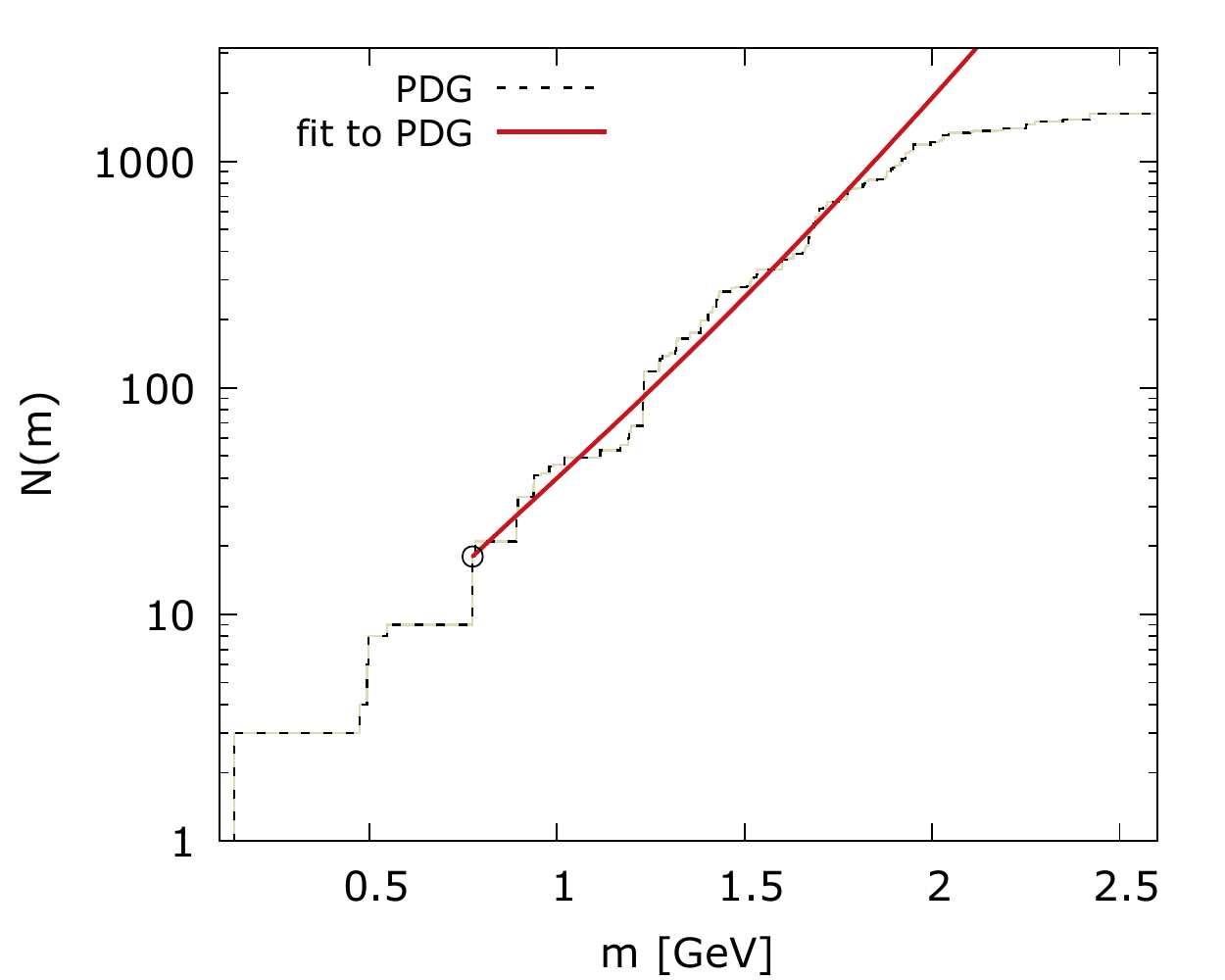}
			\label{fig:cumulant_a}}
		\centering\subfigure[]{\includegraphics[width=1\columnwidth]{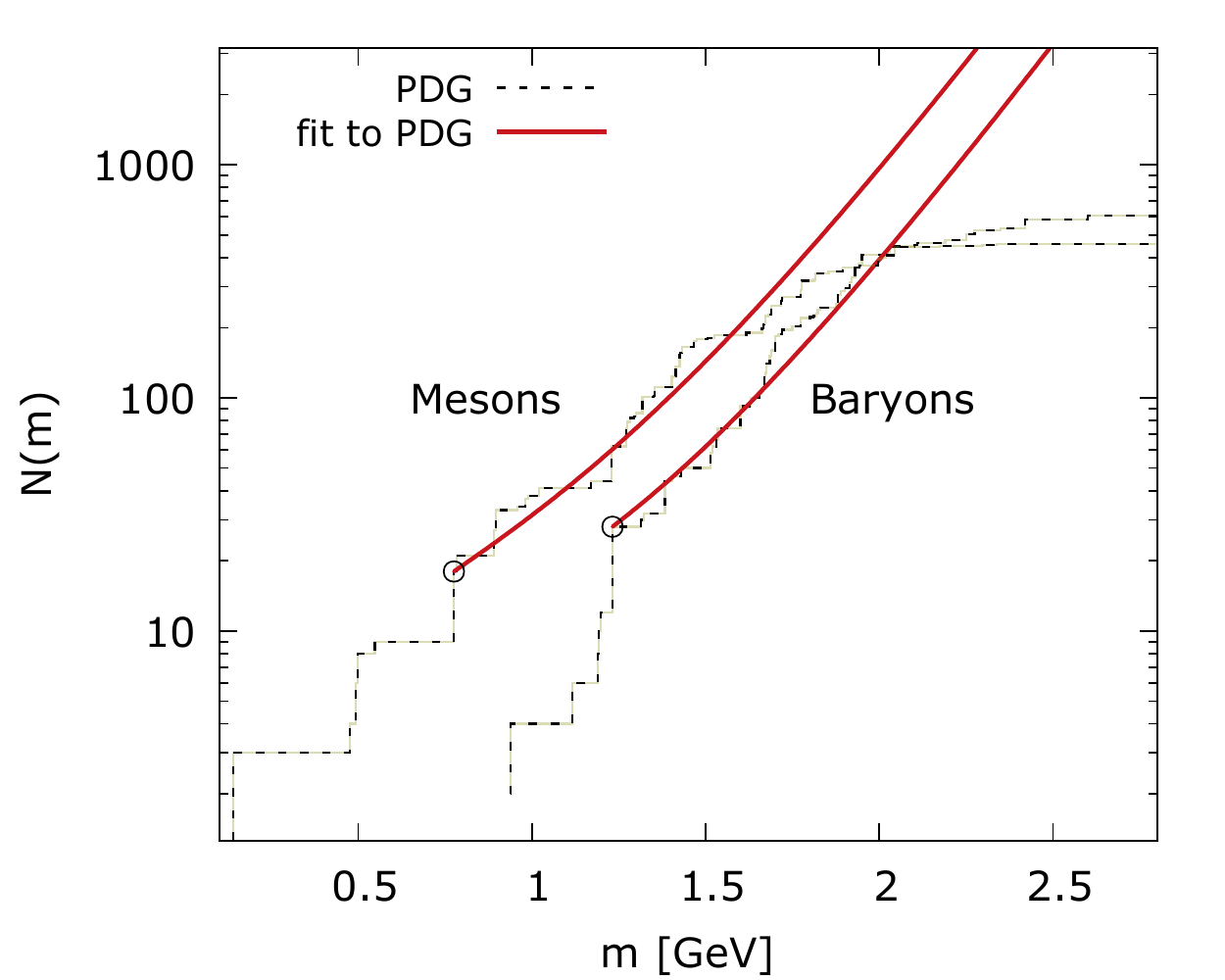}
			\label{fig:cumulant_b}}
		\centering\subfigure[]{\includegraphics[width=1\columnwidth]{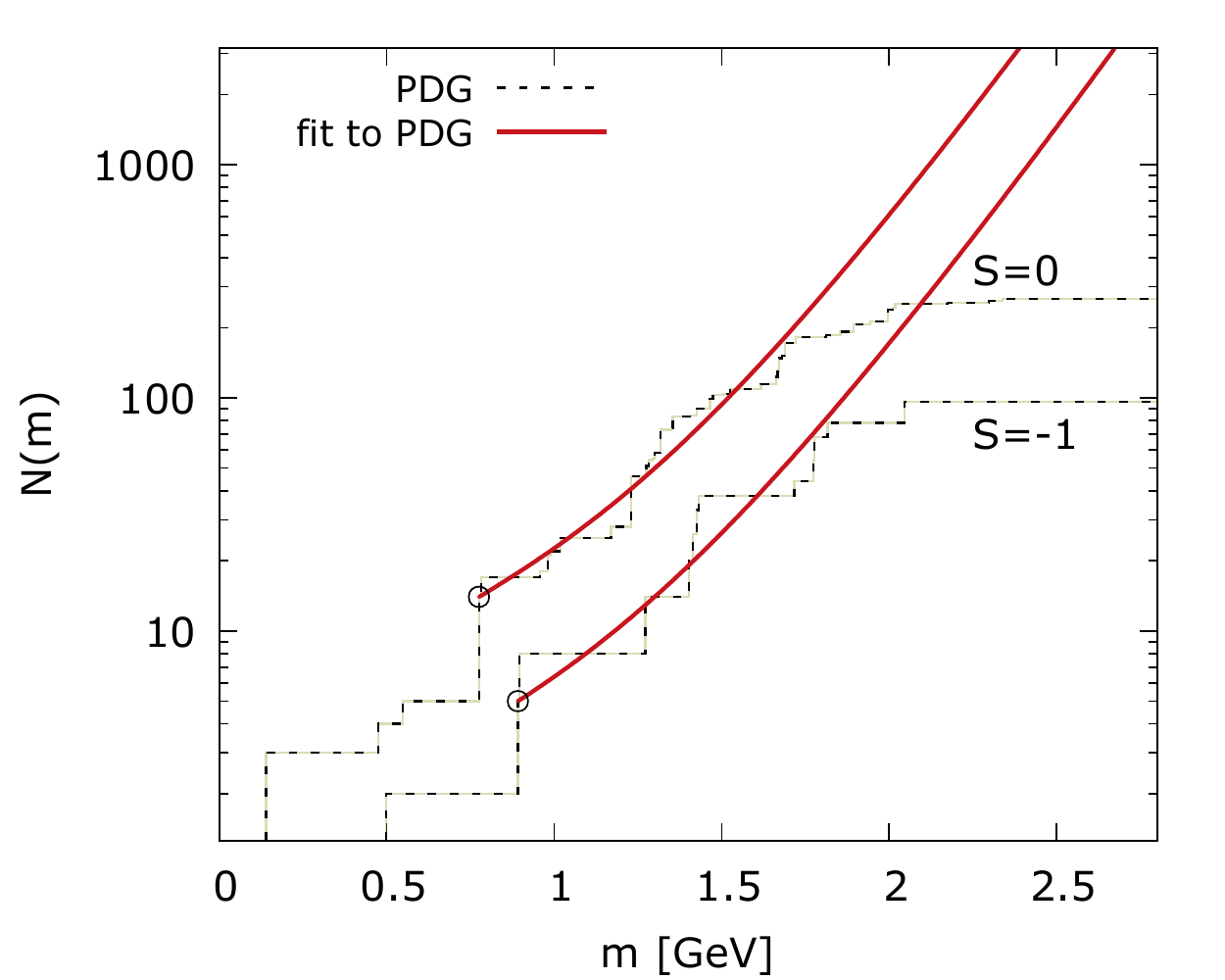}
			\label{fig:cumulant_c}}
		\centering\subfigure[]{\includegraphics[width=1\columnwidth]{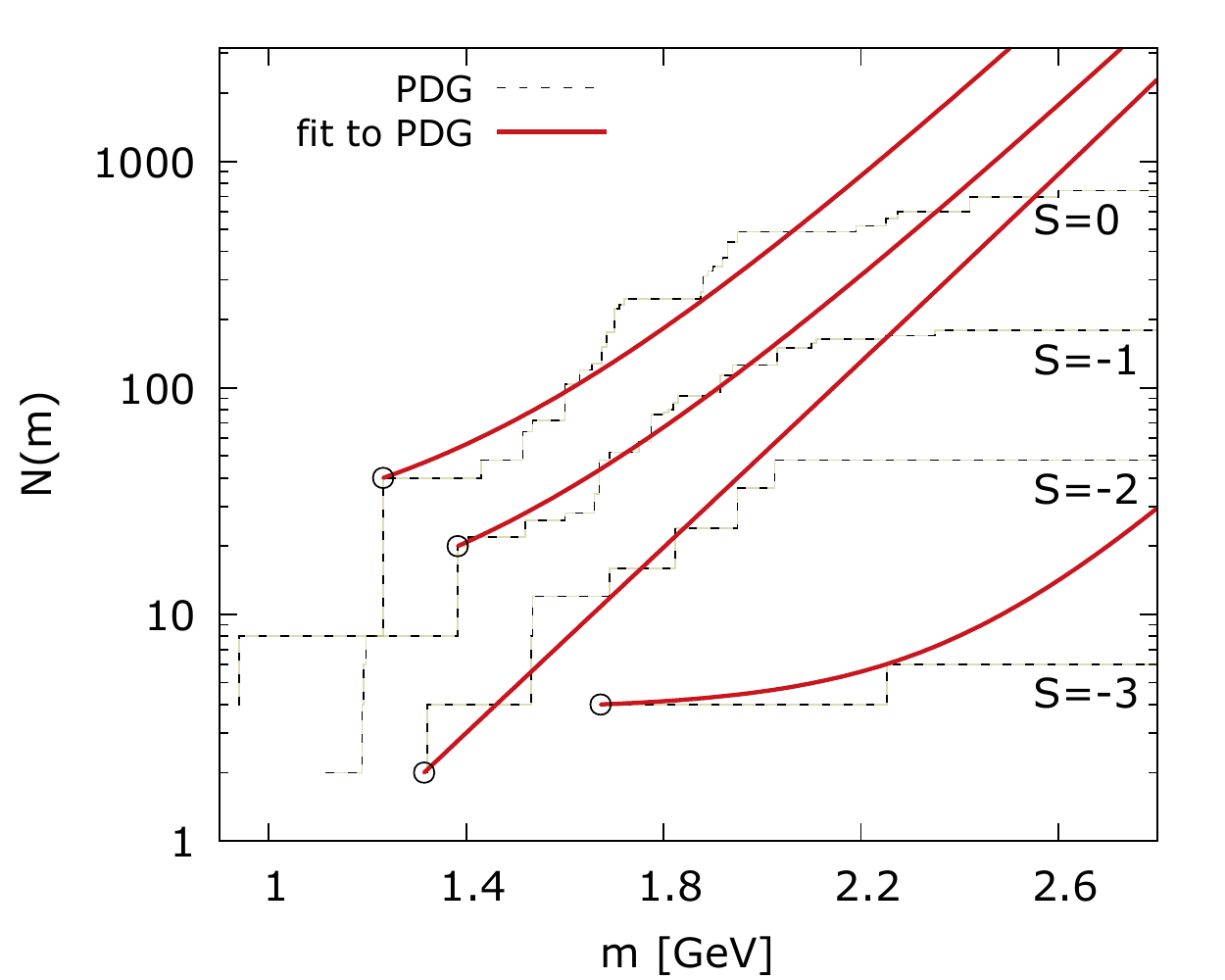}
			\label{fig:cumulant_d}}
		\caption{(Color online) Cumulants of the PDG mass spectrum in different sectors of quantum number: (a) all hadrons; (b) mesons and baryons; (c) mesons of different strangeness; (d) baryons of different strangeness. The lines are obtained from the fit of Eqs.~(\ref{eq10}) and~(\ref{eq13}) to the PDG data with the parameters listed in Table~\ref{tab:1} (see text).}
		\label{fig:cumulant}\label{fig1}
	\end{figure*}
	
	The spectrum of experimentally established hadrons, summarized by the PDG~\cite{pdg}, accounts for all identified particles and resonances, i.e., confirmed mesons and baryons granted with a three- or four-star status, of masses up to $m_M \simeq 2.4 \, {\rm GeV}$ and $m_B \simeq 2.6 \, {\rm GeV}$ respectively. The investigation of higher excited states remains a significant challenge for the experiments due to the complicated decay properties and large widths of the resonances.
	
	Instead of the hadron mass spectrum~(\ref{DEF:rho_pdg}), the medium composition can be characterized by the cumulant~\cite{Broniowski:2000bj}
	\begin{eqnarray}
	\label{DEF:cumulant_pdg}\label{eq3}
	N^{\rm{HRG}}(m) = \sum_{i} d_i \theta\left(m - m_i\right) \textrm{, }
	\end{eqnarray}
	such that
	\begin{eqnarray}\label{eq4}
	\rho^{\rm{HRG}} = {{\partial N^{\rm{HRG}}}\over { \partial  m}}.
	\end{eqnarray}
	Thus, $N^{\rm{HRG}}(m)$ counts the number of degrees of freedom with masses below $m$.
	
	Since the spectrum~(\ref{DEF:rho_pdg}) is additive in different particle species, it can be decomposed into a sum of contributions from mesons and baryons, as well as a sum of particles with definite strangeness.
	
	Figure~\ref{fig:cumulant} shows the cumulants in different sectors of hadronic quantum number with inputs from the PDG. The cumulant of all hadrons is seen in Fig.~\ref{fig:cumulant_a} to rapidly increase with mass. For $m\leq  2 \, {\rm GeV}$ such increase is almost linear, indicating that the hadron mass spectrum is exponential, as predicted by Hagedorn in the context of SBM~\cite{Hagedorn:1965st,Hagedorn:1971mc}.
	
	A rapid increase in the number of states is also seen, in Figs.~\ref{fig:cumulant_b} and~\ref{fig:cumulant_c}, to appear separately for the mesonic and baryonic sector, as well as for the strange and non-strange mesons with $m<2 \, {\rm GeV}$. Baryons of different strangeness, as illustrated in Fig.~\ref{fig:cumulant_d}, follow a similar trend with the exception of $|S|= 3$ baryons, which consists only of $\Omega$ hyperons.
	
	For an uncorrelated gas of particles (and antiparticles) with a mass spectrum $\rho(m)$, the thermodynamic pressure $\hat{P}=P/T^4$ is obtained as
	\begin{align}
	\hat{P}(T,V,\vec{\mu}) =\pm &
	\int \dd m \, \rho(m) \;
	\int   \, \frac{ \dd\hat{p}}{2\pi^2} \; \hat{p}^2 & \nonumber \\
	&\times\left[\ln (1\pm \lambda \, e^{-\hat{\epsilon}})+
	\ln (1\pm \lambda^{-1} e^{-\hat{\epsilon}}) \right]\textrm{,}
	\label{eq5}
	\end{align}
	where $\hat{p} = p/T$, $\hat{m} = m/T$, $\hat{\epsilon}=\sqrt{\hat{p}^2+\hat{ m}^2}$, and the $(\pm)$ sign refers to fermions and bosons respectively. For a particle of mass $m$, carrying baryon number $B$, strangeness $S$ and electric charge $Q$, the fugacity $\lambda$ reads
	\begin{align}
	\lambda(T,\vec\mu )= \exp \left({{B\hat{\mu}_B+S\hat{\mu}_S+Q\hat{\mu}_Q} }\right)\textrm{,}
	\label{eq6}
	\end{align}
	where $\hat{\mu} = \mu/T$. Note that, for scalar particles with vacuum quantum number, the antiparticle term should be dropped to avoid double counting.
	
	Expanding the logarithm and performing the momentum integration in Eq.~(\ref{eq5}) with the discrete mass spectrum $\rho^{\rm HRG}$ in Eq.~(\ref{eq2}), one obtains
	
	
	\begin{align}
	\hat{P}=\sum_i{{d_i}\over{2\pi^2}} & \sum_{k=1}^\infty  {{(\pm 1)^{k+1}}\over {k^2}} \, \hat{ m}_i^2 \, K_2(k\hat{m}_i) \, \lambda^k \rm ,
	\label{eq7}
	\end{align}
	
	\noindent where the first sum over $i$ includes the contributions of all known hadrons and antihadrons, and $K_2$ is the modified Bessel function. The upper and lower sign is for bosons and fermions respectively. The Boltzmann approximation corresponds to retaining only the first term in summation.
	
	
	The thermodynamic pressure in Eq.~\eqref{eq5}, through the mass spectrum $\rho$, contains all the relevant information about the distribution of mass and quantum number of the medium. Thus, it allows for the study of different thermodynamic observables, including fluctuations of conserved charges.
	
	Furthermore, one can turn this argument around, and explore the implication of the thermodynamic observables on medium composition. This approach has been applied, for example, in the pure gauge theory to extract an effective glueball mass and spectrum based on the lattice results on pressure and trace anomaly \cite{Meyer:2009tq, Caselle:2015tza}. For the case of QCD, the constraint imposed by the trace anomaly on the hadronic spectrum has been investigated \cite{Arriola:2014bfa, Arriola:2015gra}. In this work, we focus on the impact of recent LQCD data on the fluctuations of conserved charges.

	\section{Hadron Resonance Gas and LQCD}
	\label{HRG_LQCD}
	
	LQCD provides a theoretical framework to calculate the equation of state and bulk properties of strongly interacting matter at finite temperature. The first comparison of the equation of state calculated on the lattice with that derived from Eq.~(\ref{eq7}) has shown that thermodynamics of hadronic matter is well approximated by the HRG with mass spectrum generated on the corresponding lattice~\cite{tawfik1,tawfik2}.

	Presently, we have a rather detailed picture of the thermodynamics of hadronic matter, thanks to the advancement of LQCD simulations with physical quark masses and to the progress in extrapolating observables to the continuum limit~\cite{Borsanyi:2013bia,eqsf1,eqsf2}. Thus, a direct comparison of the equation of state from Eq.~(\ref{eq5}) and LQCD can be performed with the physical mass spectrum~\cite{reseqsf,Bazavov:2012jq,ratti}.
	
	\begin{figure*}[htp!]
		\centering\subfigure[]{\centering\includegraphics[width=1\columnwidth]{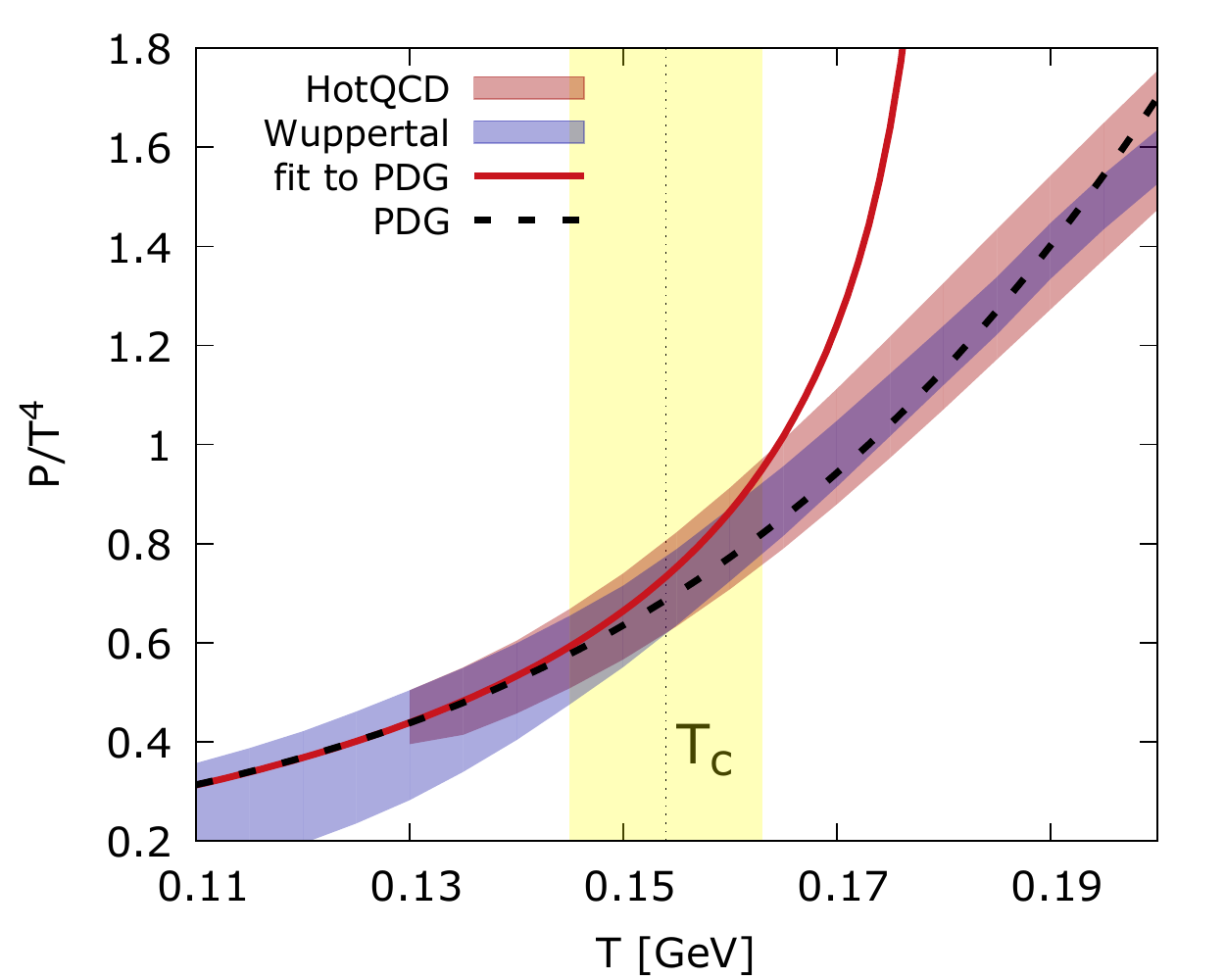}
			\label{pressure}}
		\centering
		\centering\subfigure[]{\includegraphics[width=1\columnwidth]{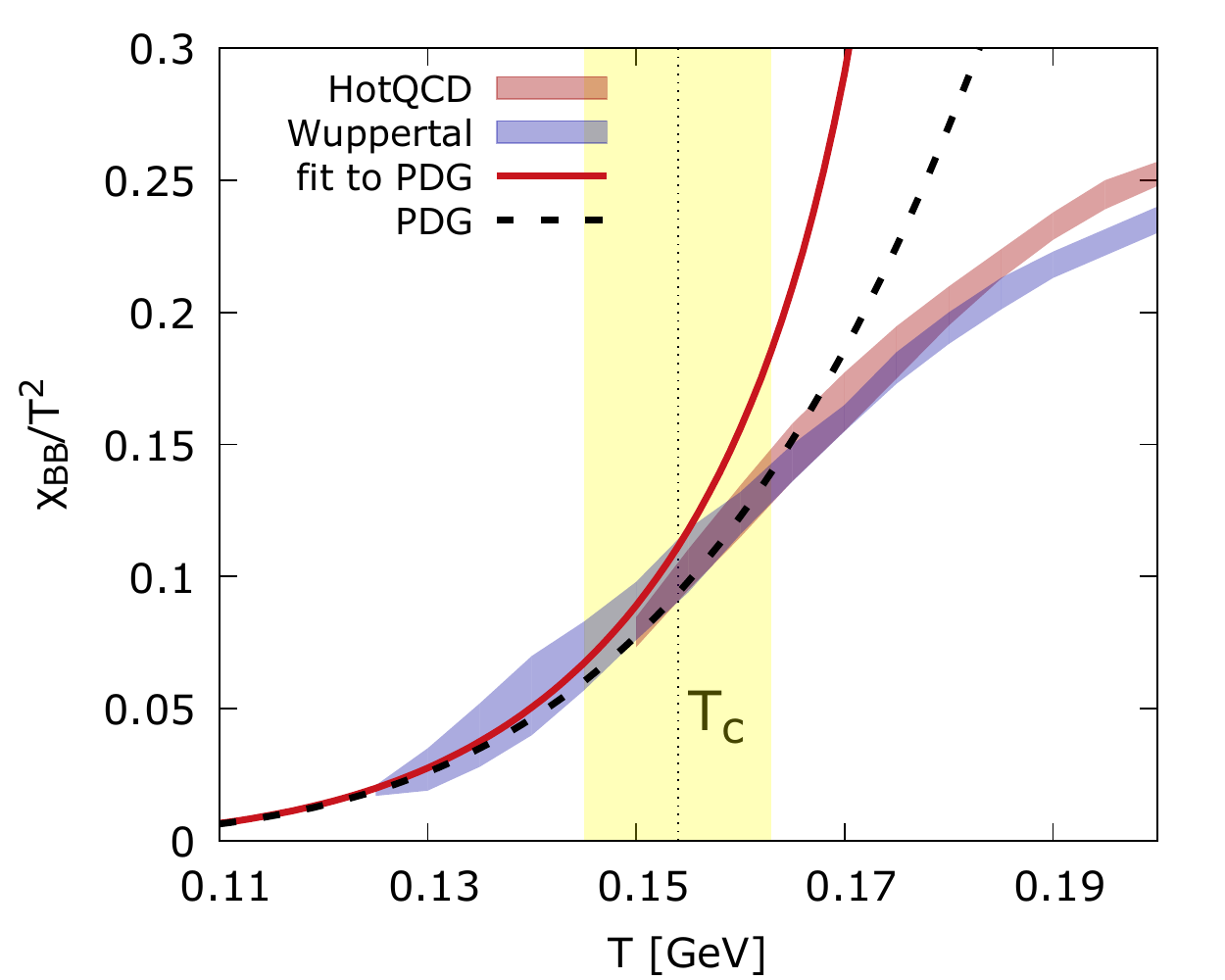}
			\label{bb}}
		\caption{(Color online) Lattice QCD results of HotQCD~\cite{eqsf1,Bazavov:2012jq} and Budapest-Wuppertal Collaborations~\cite{Borsanyi:2013bia,Borsanyi:2011sw} for different observables in dimensionless units: (a) the thermodynamic pressure; (b) the net-baryon number fluctuations $\hat{\chi}_{\rm BB}$. Also shown are the HRG results for the discrete PDG mass spectrum (dashed line) and for the effective mass spectrum in Eq.~(\ref{eq12}), which contains a continuous part to describe the effects of massive resonances (continuous line).}
		\label{pressure_bb}
	\end{figure*}
	
	In Fig.~\ref{pressure} we show the temperature dependence of the thermodynamic pressure obtained recently in lattice simulations with physical quark masses~\cite{eqsf1,Borsanyi:2013bia}. The bands of the LQCD result indicate the systematic errors due to continuum extrapolation. The vertical band marks the temperature $T_c = 155(1)(8) \, {\rm MeV}$, which is the chiral crossover temperature from the hadronic phase to the quark-gluon plasma~\cite{tc}. These LQCD results are compared in Fig.~\ref{pressure} with predictions of the HRG model for the mass spectrum~(\ref{DEF:rho_pdg}), which includes all known hadrons and resonances listed by the PDG~\cite{pdg}.
	
	There is a clear coincidence of the HRG and LQCD results on the equation of state at low temperatures. The pressure is strongly increasing with temperature towards the chiral crossover. This behavior is well understood within HRG as the consequence of growing contributions from the escalating number of higher resonances.
	
	Although HRG formulated with a discrete mass spectrum does not exhibit any critical behavior, it nevertheless reproduces remarkably well the lattice results in the hadronic phase. This agreement has now been extended to the fluctuations and correlations of conserved charges~\cite{Bazavov:2012jq,Borsanyi:2011sw,ejiri1,ejiri2}.
	
	In a thermal medium, the second-order fluctuations and correlations of conserved charges are quantified by the generalized susceptibilities
	\begin{eqnarray}
	\label{sus1}\label{eq8}
	\hat{ \chi}_{xy} = \frac{\partial^{2}\hat P}{\partial \hat \mu_x\partial \hat \mu_y}  \textrm{,}
	\end{eqnarray}
	where $(x,y)$ are conserved charges, which in the following are restricted to the baryon number $B$ and strangeness $S$.
	
	For HRG with a discrete mass spectrum of Boltzmann particles, $\hat{\chi}_{xy}$ is obtained from Eq.~(\ref{eq7}) as
	\begin{eqnarray}
	\label{eq9}
	\hat{ \chi}^{\rm HRG}_{xy}\Big|_{\hat \mu_x = \hat\mu_y=0} = \frac{1}{\pi^2}\sum_{i} d_i {\hat{m}_i^2}K_2\left({\hat{m}_i}\right)x_iy_i \textrm{.}
	\end{eqnarray}
	
	The susceptibilities~(\ref{sus1}) and in particular~(\ref{eq9}) are observables sensitive to the quantum numbers of medium constituents. Thus, $\hat\chi_{xy}$ can be used to identify contributions of different particle species to QCD thermodynamics~\cite{ejiri1,ejiri2}.
	
	\begin{figure*}[htp!]
		\centering\subfigure[]{\centering\includegraphics[width=1\columnwidth]{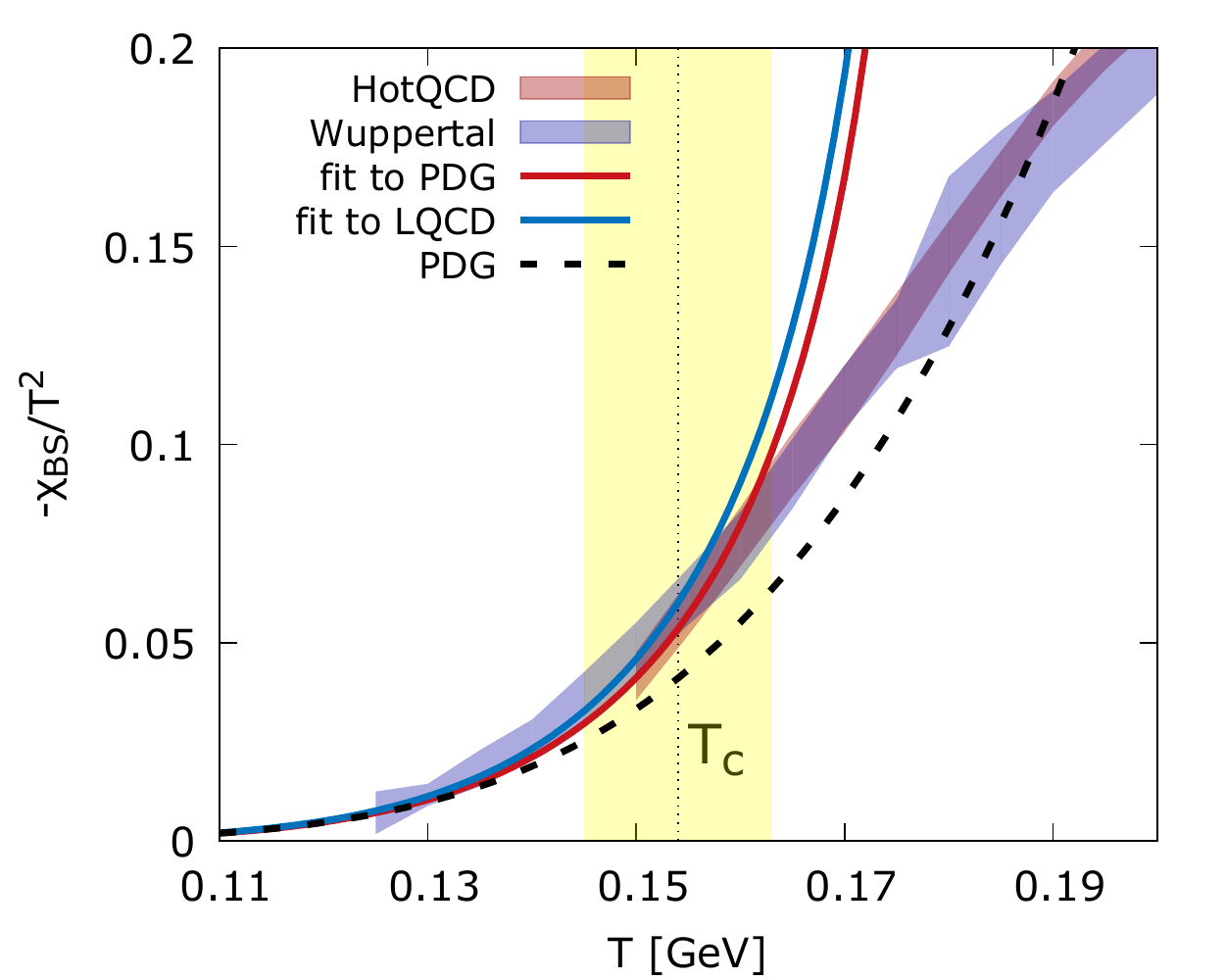}
			\label{bs}}
		\centering\subfigure[]{\centering\includegraphics[width=1\columnwidth]{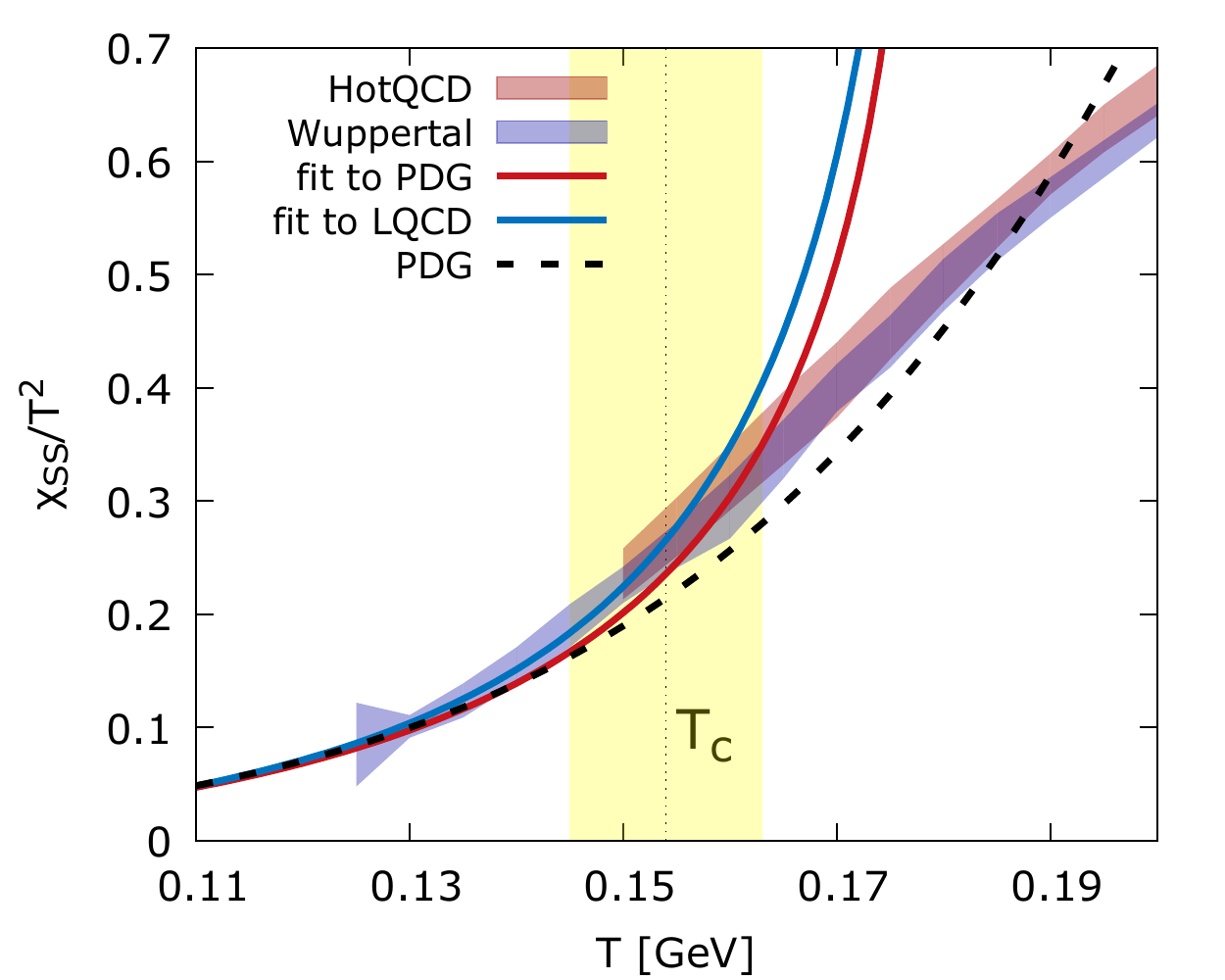}
			\label{ss}}
		\caption{ (Color online) As in Fig.~\ref{pressure_bb}, but for baryon-strangeness correlations $\hat{\chi}_{\rm BS}$ (a) and for strangeness fluctuations $\hat{\chi}_{\rm SS}$ (b). Also shown are the corresponding results obtained from the least-square fits to lattice data up to $T\simeq 156 \, {\rm MeV}$.}
		\label{bs_ss}
	\end{figure*}
	
	Recent LQCD calculations of the HotQCD Collaboration~\cite{Bazavov:2012jq} and Budapest-Wuppertal Collaboration~\cite{Borsanyi:2013bia, Borsanyi:2011sw} provide results on different fluctuations and correlations of conserved charges. Thus, the apparent agreement of HRG and LQCD, seen on the level of the equation of state, can be further tested within different hadronic sectors~\cite{Bazavov:2012jq}.
	
	In Figs.~\ref{bb} and~\ref{bs_ss} we show the LQCD results on the fluctuation of net-baryon number, strangeness, as well as the baryon-strangeness correlations. They are compared to the HRG model, formulated with the PDG mass spectrum. From Fig.~\ref{bb}, it is clear that the net-baryon number fluctuation in the hadronic phase are well described by HRG, whereas strangeness fluctuation $\hat\chi_{\rm SS}$ in Fig.~\ref{ss} and $\hat\chi_{\rm BS}$ correlation in Fig.~\ref{bs} are underestimated in the low-temperature phase.
	
	Following an analysis of the relations between different susceptibilities of conserved charges, it was argued in Ref.~\cite{Bazavov:2012jq} that deviations seen in Fig.~\ref{bs} can be attributed to the missing resonances in the strange-baryonic sector. In view of Fig.~\ref{ss}, similar conclusion can be drawn for the strange mesons.
	
	In general, the contributions of heavy resonances in HRG are suppressed due to the Boltzmann factor. However, the relative importance of these states depends on the observable. In the hadronic phase, the pressure is dominated by the low-lying particles. At temperature $T=150 \, {\rm MeV}$, the contribution to the pressure from particles and resonances with mass $M>1.5 \, {\rm GeV}$ is of the order of 7$\%$. However, in the fluctuations of the net-baryon number and baryon-strangeness correlations, such a contribution is already significant and amounts to 26$\%$ and 33$\%$, respectively.
	
	Contributions from missing heavy states could be the potential origin of the observed differences between LQCD results and HRG predictions on fluctuations and correlations of conserved charges in the strange sector, shown in Figs.~\ref{bs} and~\ref{ss}. 
	
	\section{Hagedorn mass spectrum and LQCD fluctuations}
	\label{Hagedorn_LQCD}
	
	\subsection{The Hagedorn mass spectrum}
	
	To account for the unknown resonance states at large masses we adopt the continuous Hagedorn mass spectrum with the parametrization
	\begin{eqnarray}
	\label{DEF:rho_hagedorn}\label{eq10}
	\rho^H(m) = \frac{a_0\;}{\left(m^2+m_0^2\right)^{5/4}} e^{m/T_H}\textrm{,}
	\end{eqnarray}
	and its corresponding cumulant
	\begin{eqnarray}
	\label{DEF:cumulant_hag}\label{eq11}
	N^H(m) = \int\limits_0^m \dd m' \; \rho^H(m') \textrm{,}
	\end{eqnarray}
	where $T_H$ is the Hagedorn limiting temperature, whereas $a_0$ and $m_0$ are additional free parameters.
	
	In general, the parameters of $\rho(m)$ can be calculated within a model, e.g., in SBM~\cite{Hagedorn:1971mc,Frautschi:1971ij}. In the following, we constrain the Hagedorn temperature and the weight parameters $(a_0,m_0)$ in Eq.~(\ref{eq10}) based on the mass spectrum of the PDG and the lattice data. In addition, we assume that the same exponential functional form holds separately for hadrons in different sectors of quantum number, i.e., for mesons and baryons with or without strangeness.
	
	\begin{table*}[htp!]
		\begin{tabular}{|l||c|c||c|c|}
			\hline
			& \multicolumn{2}{|c||}{Fit to PDG}							& \multicolumn{2}{|c|}{Fit to LQCD}							\\ \hline
			& $m_0$ [${\rm GeV}$]	& $a_0(m_0)$ [${\rm GeV}^{3/2}$]	& $m_0$ [{\rm GeV}]		& $a_0(m_0)$ [${\rm GeV}^{3/2}$]	\\ \hline\hline
			$\rho_H$			& 0.529(22)				& 0.744(40)								& 0.425(24)			& 0.573(36)							\\ \hline
			$\rho_{B}$			& 0.145(23)				& 0.135(7)								& 0.078(13)			& 0.108(9)							\\ \hline
			$\rho_{B}^{S=0}$	& 0.053(8)				& 0.064(12)								&					&									\\ \hline
			$\rho_{B}^{S=-1}$	& 0.051(12)				& 0.046(6)								& 0.193(96)(122)	& 0.067(27)							\\ \hline
			$\rho_{B}^{S=-2}$	& 1.453(441)			& 0.023(20)								& 2.469(456)(297)	& 0.091(47)							\\ \hline
			$\rho_{B}^{S=-3}$	& 0.00194(0)			& 0.00027(0)							&					&									\\ \hline
			$\rho_{M}$			& 0.244(17)				& 0.341(19)								&					&									\\ \hline
			$\rho_{M}^{S=0}$	& 0.183(19)				& 0.212(17)								&					&									\\ \hline
			$\rho_{M}^{S=-1}$	& 0.183(43)				& 0.060(9)								& 0.378(32)(95)		& 0.099(24)							\\ \hline
		\end{tabular}\qquad\qquad\qquad
		\begin{tabular}{|l||c|c|}
			\hline
			& \multicolumn{2}{|c|}{Constraint}				\\ \hline
			& $m_x$ [${\rm GeV}$]	& $N^{\textrm{HRG}}(m_x)$	\\ \hline\hline
			$\rho_H$			& 0.77526				& 18						\\ \hline
			$\rho_{B}$			& 1.2320				& 28						\\ \hline
			$\rho_{B}^{S=0}$	& 1.2320				& 40						\\ \hline
			$\rho_{B}^{S=-1}$	& 1.3828				& 20						\\ \hline
			$\rho_{B}^{S=-2}$	& 1.31486				& 2							\\ \hline
			$\rho_{B}^{S=-3}$	& 1.67245				& 4							\\ \hline
			$\rho_{M}$			& 0.77526				& 18						\\ \hline
			$\rho_{M}^{S=0}$	& 0.77526				& 14						\\ \hline
			$\rho_{M}^{S=-1}$	& 0.89166				& 5							\\ \hline
		\end{tabular}
		\caption{
			(Left) Parameters of the Hagedorn mass spectra in Eqs.~(\ref{eq10}) and (\ref{eq12}), in different sectors, obtained from fits to PDG and LQCD data. The Hagedorn temperature has been set to $T_H = 180 \, {\rm MeV}$. Sectors of all hadrons, all mesons, and nonstrange mesons include both the particles' and antiparticles' contributions.  In matching the LQCD results, the data for pressure and second-order fluctuations are compared with the HRG model through Eqs.~(\ref{eq5}) and (\ref{DEF:HAG_fluct}). Also shown are the errors of $m_0$ arising from the least-square fits, which induce the uncertainties in $a_0(m_0)$ through Eq.~(\ref{DEF:constraint2}). For the sectors $\rho_{B}^{S=-1}$, $\rho_{B}^{S=-2}$, and $\rho_{M}^{S=-1}$, the systematic errors in the approximation schemes are also included (see text). (Right) The constraint on the continuous mass spectrum in each sector, given in Eq.~(\ref{DEF:constraint}). 
		}
		\label{tab:1}
	\end{table*}
	
	The analysis of the experimental hadron spectrum, in the context of Hagedorn exponential form, has been extensively discussed in the literature~\cite{Hagedorn:1971mc,raf1,raf2,Majumder:2010ik}. In one of the recent studies~\cite{Broniowski:2000bj,Broniowski:2004yh}, it was shown that, in fitting the Hagedorn spectrum to experimental data, one arrives at different limiting temperatures for mesons, baryons, and hadrons with different electric charges. In particular, with $\rho(m)$ from Eq.~(\ref{eq10}), the limiting temperature for mesons, $T_H^M\simeq 195  \, {\rm MeV}$, was extracted to be considerably larger than that for baryons, $T_H^B\simeq 140 \, {\rm MeV}$. Such Hagedorn limiting temperatures, however, are inconsistent with recent lattice results, which show that the changes from the hadronic to the quarks and gluons degrees of freedom in different sectors appear in the same narrow temperature range of the chiral crossover. Thus, the Hagedorn temperature of baryons should appear beyond the chiral crossover, i.e., $T_H^B>155 \, {\rm MeV}$. In addition, the LQCD data on $\hat\chi_{\rm BB}$ are consistent with the discrete PDG baryon mass spectrum up to $T\simeq 160 \, {\rm MeV}$. This seems to suggest that large contributions form heavy resonances are not expected in $\hat\chi_{\rm BB}$ at $T_H^B<155 \, {\rm MeV}$.
	
	From the above one concludes that it is very unlikely for the Hagedorn limiting temperatures in various hadronic sectors to differ substantially. Moreover, they are expected to be larger than the chiral crossover temperature. Consequently, the extracted Hagedorn temperature $T_H^B\simeq 140 \, {\rm MeV}$ for baryons in Refs.~\cite{Broniowski:2000bj,Broniowski:2004yh}, though mathematically correct, is disfavored by LQCD.
	
	The reason for the very different Hagedorn temperatures for mesons and baryons is that the extraction of the parameters in Eq.~(\ref{eq10}) has been performed over the whole mass range of the PDG data. The low-lying baryons drives the fit towards a lower $T_H$, resulting in the deviation of Hagedorn temperatures among different sectors.

	To avoid the above problem, we adopt Hagedorn's idea to treat the contributions of ground state particles\footnote{Particles that do not decay under the strong interaction. In this context, there are no ground states in the $|S|=2$ and $|S|=3$ baryonic sectors.} separately from the exponential mass spectrum. In addition, we start the continuous part of the spectrum from the onset of the first resonance in the corresponding sector. Therefore, we apply the following mass spectrum
	\begin{eqnarray}
	\label{DEF:rho_mix}\label{eq12}
	\rho(m) = \sum_{i} d_i\delta(m - m_i)+ \rho^H(m) \theta(m - m_x) \textrm{,}
	\end{eqnarray}
	and the corresponding cumulant
	\small
	\begin{eqnarray}
	\label{DEF:cumulant_mix}\label{eq13}
	N(m) = \sum_{i} d_i\theta(m - m_i)+ \theta(m - m_x)\int\limits_{m_x}^m\dd m\;\rho^H(m) \textrm{,}
	\end{eqnarray}
	\normalsize
	\noindent where $\rho^H(m)$ is given by Eq.~(\ref{eq10}). The index $i$ counts the hadronic ground states, i.e., states with masses less than $m_x$ of the first resonance in the corresponding channel. 
	
	 With such a prescription for analyzing the hadronic spectrum, we find no practical advantage in treating the continuous $\rho^H(m)$ as a two-parameter function of ($m_0$, $a_0$). We therefore impose the following constraint on the continuous mass spectrum
		
		\begin{eqnarray}
		\label{DEF:constraint}
		N^{\textrm{HRG}}(m_x) = \int\limits_0^{m_x}\dd m\; \rho^H(m).
		\end{eqnarray}
		
		\noindent This way, $\rho^H$ is reduced to a function of a single parameter $m_0$. The parameter $a_0$ can be determined by
		
		\begin{eqnarray}
		\label{DEF:constraint2}
		a_0(m_0) = N^{\textrm{HRG}}(m_x) \left[\int\limits_0^{m_x}\dd m\; \frac{e^{m/T_H}}{{\left(m^2+m_0^2\right)^{5/4}}}\right]^{-1} \textrm{.}
		\end{eqnarray}

	The above spectrum can now be compared with the experimental data listed by the PDG, in different sectors of quantum number. From the analysis of the mass spectrum parameters of all hadrons, we find that the best description is obtained with $T_H\simeq 180 \, {\rm MeV}$. This value is consistent with that recently found in Ref.~\cite{Majumder:2010ik}. In addition, $T_H\simeq 180 \, {\rm MeV}$ is the largest temperature obtained as the solution of the Bootstrap equation~\cite{satz}. In the following, we apply the same $T_H$ for strange and nonstrange hadrons.

	In Fig.~\ref{fig:cumulant} we show that the spectra of PDG hadrons in different sectors are indeed properly described by the asymptotic mass spectrum~(\ref{eq12}) with a common Hagedorn temperature $T_H\simeq 180 \, {\rm MeV}$. The weight parameters $(a_0,m_0)$ in Eq.~(\ref{eq10}) are determined by the composition and decay properties of the resonances, hence, they are distinct for each hadronic quantum number. The optimal sets of parameters of $\rho(m)$ in Eq.~(\ref{eq10}) are summarized in Table~\ref{tab:1}. The corresponding mass spectra are shown in Fig.~\ref{fig:cumulant} as continuous lines, whereas circles indicate the lowest masses $m_x$ of the corresponding fit. Also shown, as broken lines in Fig.~\ref{fig:cumulant}, are the extrapolated cumulants below $m_x$.

	\begin{figure*}[htp!]
		\centering\subfigure[]{\includegraphics[width=1\columnwidth]{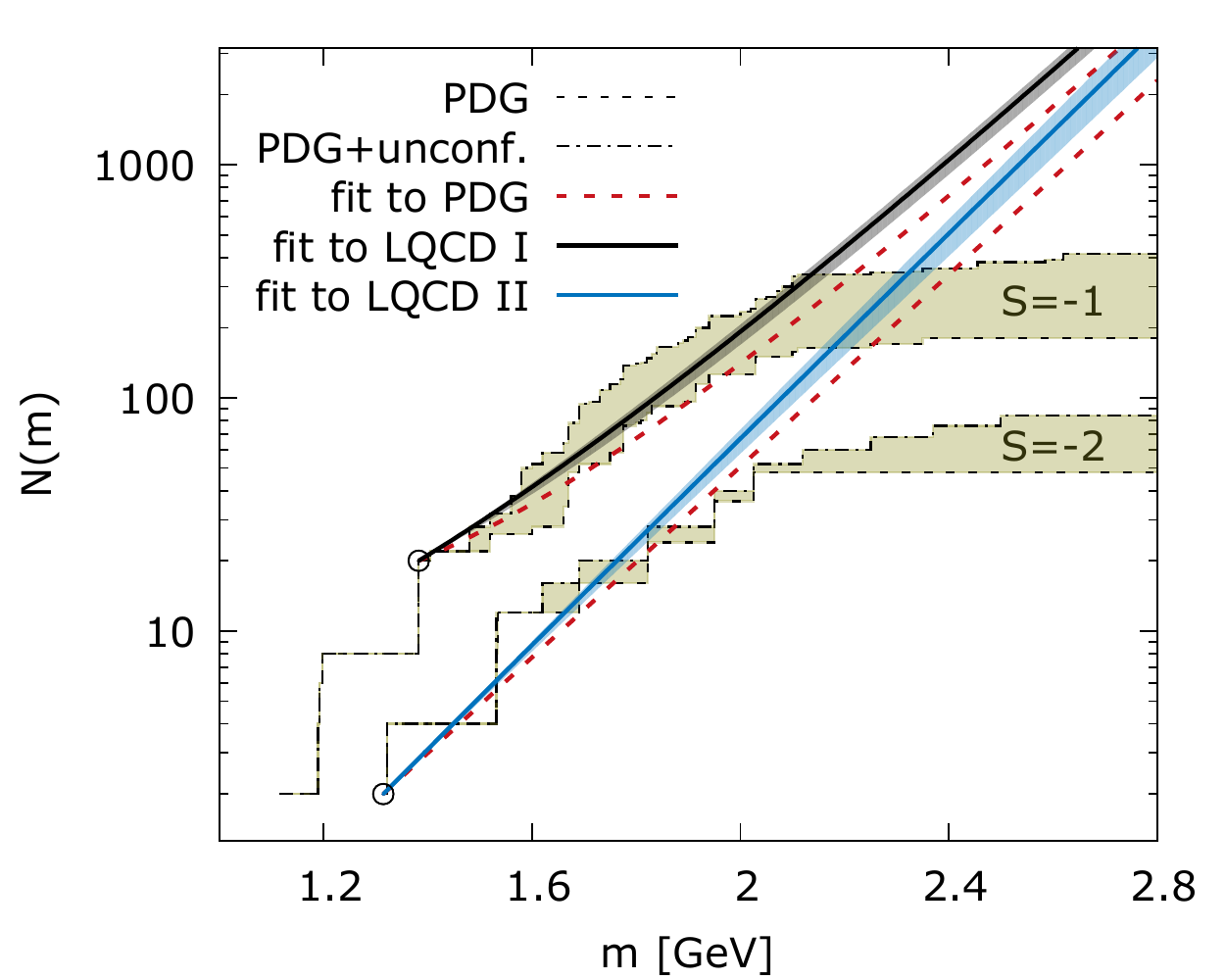}
			\label{figb}}
		\centering\subfigure[]{\includegraphics[width=1\columnwidth]{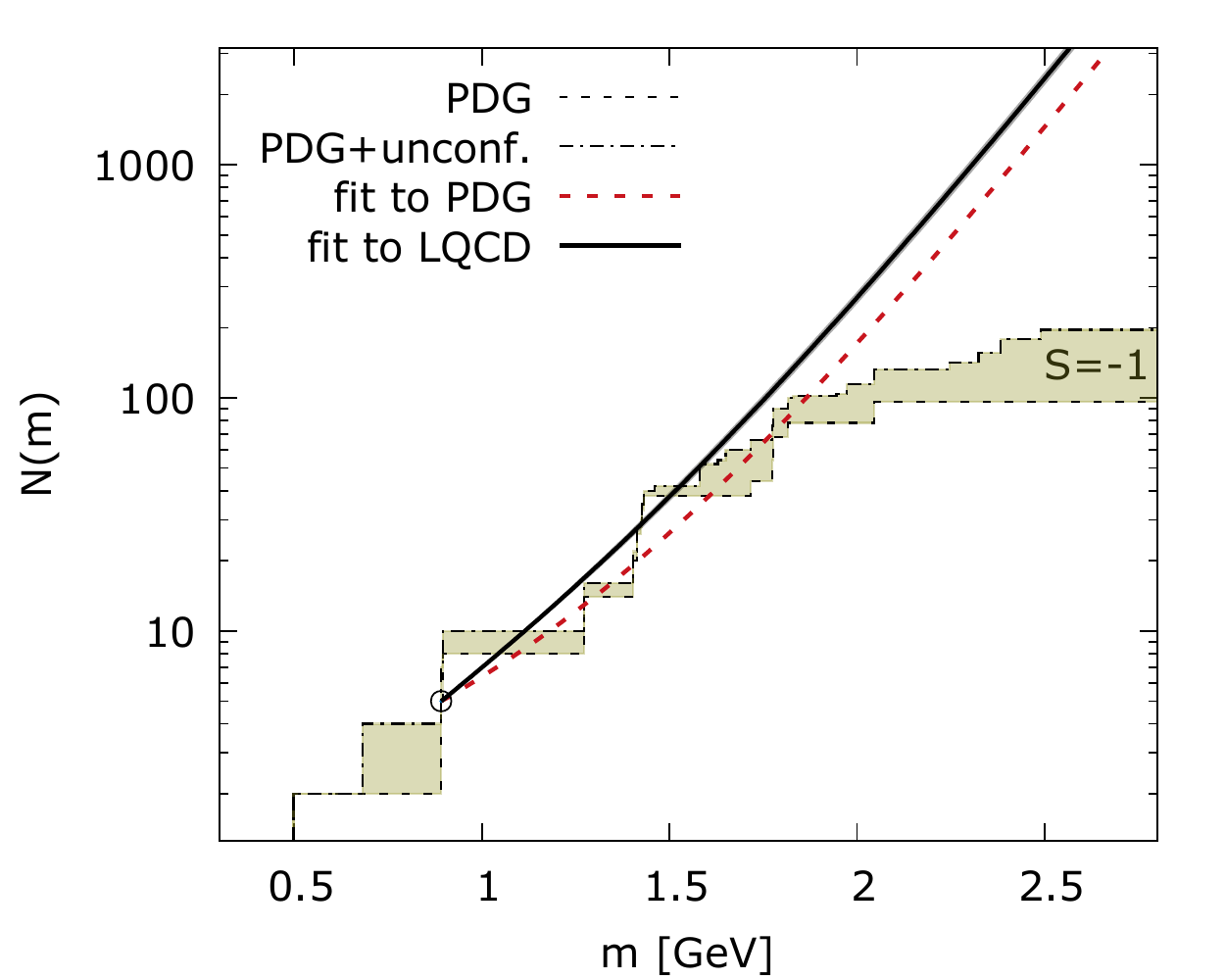}
			\label{figm}}
			\caption{(Color online) Cumulants of the discrete PDG mass spectrum (black dashed line) and the corresponding fits (red dashed line) for (a) strange baryons and (b) strange mesons. Also shown are the cumulants containing in addition the unconfirmed states (broken-dashed line). Continuous lines are obtained by matching the LQCD results to the continuous mass spectra through Eq.~(\ref{DEF:HAG_fluct}), assuming that the missing strange baryons come solely from the $|S|=1$ sector $\left[ \textrm{scheme (I)}\right]$ or $|S|=2$ sector $\left[ \textrm{scheme (II)}\right]$ (see text).}
		\label{fig:cumulant_fit}
	\end{figure*}
	
	It is important for the decomposition of the hadron mass spectrum (\ref{eq12}) into different sectors, using parameters from Table~\ref{tab:1}, to produce results that are thermodynamically consistent. Thus, e.g., the total pressure $\hat P^H$ obtained from Eq.~(\ref{eq5}) with the mass spectrum from Fig.~\ref{fig:cumulant_a} should be consistent with the sum of meson $\hat P_M$ and baryon $\hat P_B$ pressures, calculated with the mass spectra in Fig.~\ref{fig:cumulant_b}. Similar results should hold for the pressure when adding up the contributions from strange particles in different sectors. This consistency check provides further constraints on the mass spectrum parameters presented in Table~\ref{tab:1}.

	With the PDG mass spectrum extrapolated to the continuum, we can now test whether heavy resonances can reduce or eliminate the discrepancies between HRG and LQCD on baryon-strangeness correlations and strangeness fluctuations, seen in Figs.~\ref{bs} and~\ref{ss}. 
	
	The second order cumulants $\hat\chi_{xy}$, at vanishing chemical potential, are obtained in HRG as
	\small
	\begin{subequations}
		\label{DEF:HAG_fluct}
		\begin{align}
		\hat \chi^{H}_{\rm BB} &= \int\limits_0^\infty \frac{\dd m}{\pi^2} \; \rho_{{B}}(m) \hat m^2K_2\left(\hat m\right) \textrm{,}\\
		\hat \chi^{H}_{\rm SS} &=  \int\limits_0^\infty \frac{\dd m}{\pi^2} \;\left[\rho_{{M}}^{S=-1}(m) + \sum_{k=1}^{3}k^2\rho_{{B}}^{S=-k}(m)\right]\hat m^2K_2\left(\hat m\right)\textrm{,}\\
		\hat \chi^{H}_{\rm BS} &= - \int\limits_0^\infty \frac{\dd m}{\pi^2} \;\left[\sum_{k=1}^{3}k\rho_{{B}}^{S=-k}(m)\right]\hat m^2K_2\left(\hat m\right)\textrm{, }
		\end{align}
	\end{subequations}
	\normalsize
	using the mass spectrum $\rho(m)$ in Eq.~(\ref{eq12}) and the parameters presented in Table~\ref{tab:1}. 
	
	In Figs.~\ref{pressure_bb} and~\ref{bs_ss}, we show the contribution of the continuous Hagedorn mass spectrum to the pressure and different charge susceptibilities. The difference between the full line (fit to PDG) and the dashed line (PDG) comes from the inclusion of heavy resonances. The results in Figs.~\ref{bs} and~\ref{ss} indicate that heavy resonances can capture, to a large extent, the differences between HRG and LQCD for strangeness and baryon-strangeness correlations. However, at low temperatures, $\hat \chi_{\rm BS}$ still differs from the lattice. These deviations suggest that there are additional missing resonances in the PDG data in the mass range $m<2 \, {\rm GeV}$, as they begin to contribute substantially to $\hat\chi_{\rm BS}$ and $\hat\chi_{\rm SS}$ at lower temperatures.
	
	\subsection{Spectra of strange hadrons from LQCD}
	
		To identify the missing strange resonances in the Hagedorn mass spectrum, we use the LQCD susceptibility data as input for Eq.~(\ref{DEF:HAG_fluct}) to constrain $\rho(m)$ in different sectors. 
		
		We begin with the strange-baryonic sector. The $\hat\chi_{\rm BS}$ data alone do not allow for a unique determination of the contribution from a particular sector. This is because the observable depends only on a linear combination of the spectra, namely $\rho_B^S = \rho_B^{S=-1} + 2\rho_B^{S=-2} + 3\rho_B^{S=-3}$. In principle, this problem could be resolved with additional lattice data on higher-order strangeness fluctuation, e.g.,  $\hat\chi_{{\rm BBSS}}$ and the kurtosis. 
		
		For our purpose of analyzing the present data, we instead make the following simplification. The $|S|=3$ sector is restricted to those states listed by the PDG. We then make the assumption that the additional strange baryons come solely from the $|S|=1$ $\left[ \textrm{hereafter named scheme (I)}\right]$ or $|S|=2$ sector $\left[ \textrm{scheme (II)}\right]$, and treat the remaining one with the spectrum fitted to the PDG.\footnote{The errors induced by the use of the PDG-fitted spectra are introduced as systematic errors for the spectrum parameters. See Table~\ref{tab:1}.} The resulting spectrum parameters for both schemes are presented in Table~\ref{tab:1}.
		
		In Fig.~\ref{figb} we show the cumulants of the lattice-induced $\rho(m)$ under both schemes, together with the experimental spectra including the unconfirmed states from the PDG. The mass spectrum extracted with scheme $({\rm I})$ is seen in Fig.~\ref{figb} to follow the trend of the unconfirmed states of the PDG. This is not the case for scheme $({\rm II})$. Hence, the extra PDG data for the unconfirmed hyperons support the former scenario. 
		
		The $\hat \chi_{\rm SS}$ fluctuation in general receives contributions from both the strange mesons and the strange baryons. However, due to Boltzmann suppression, we expect the observable to be dominated by the mesonic contribution. This can be inferred from the fact that $\hat\chi_{\rm SS} \gg \hat \chi_{\rm BS}$. To be definite, we fix the strange baryon contribution to $\hat \chi_{\rm SS}$ by the scenario dictated by scheme $({\rm I})$. The lattice data on $\hat\chi_{\rm SS}$ then directly determines the strange-mesonic spectrum. The resulting parameter is presented in Table~\ref{tab:1}, with the corresponding cumulant shown in Fig.~\ref{figm}. 
		
		Similar to the case of strange baryons, the spectrum determined from LQCD requires additional states beyond the known strange mesons. From Fig.~\ref{figm}, we find that it exceeds even the trend set by the inclusion of the unconfirmed resonances. This may point to the existence of some uncharted strange mesons in the intermediate mass range. 

		In addition, the general conclusion of an enhanced \mbox{lattice-motivated} strange spectra, relative to the PDG, does not depend on the chosen functional form of the continuous spectrum \eqref{DEF:rho_hagedorn}. This follows from the observation that lattice data show a stronger interaction strength in the strange sector than that expected from a free gas of known hadrons. Within the framework of HRG this implies an increase in the corresponding particle content.
		
		Nevertheless, it is important to bear in mind that such conclusion, based on an ideal resonance-formation treatment of the hadron gas, is not definitive. For example, the contribution to $\hat \chi_{\rm SS}$ from the non-strange sector is also possible through the vacuum fluctuation of $s \bar{s}$ mesons. Such an effect is neglected in the current model and a theoretical investigation is under way.
	
	\section{Summary and conclusions}
	\label{conclusions}
	
	Modeling the hadronic phase of QCD by the hadron resonance gas (HRG), we have examined the contribution of heavy resonances, through the exponential Hagedorn mass spectrum \mbox{$\rho(m)\simeq m^a e^{m/T_H}$}, to the fluctuation of conserved charges. A quantitative comparison between model predictions and lattice QCD (LQCD) calculations is made, with a special focus on strangeness fluctuations and baryon-strangeness correlations.
	
	We have reanalyzed the mass spectrum of all known hadrons and resonances listed in the Particle Data Group (PDG) database. A common Hagedorn temperature, $T_H\simeq 180 \, {\rm MeV}$, is employed to describe hadron mass spectra in different sectors of quantum number. This value of $T_H$ exceeds the LQCD chiral crossover temperature $T_c = 155(1)(8) \, {\rm MeV}$. The latter signifies the conversion of the hadronic medium into a quark-gluon plasma.
	
	Applying the continuum-extended mass spectrum calculated from the PDG data, we have shown that the Hagedorn asymptotic states can partly remove the disparities with lattice results in the strange sector. 
	
	To fully identify the missing hadronic states, we perform a matching of LQCD data on strangeness fluctuations and baryon-strangeness correlations with HRG. The parameters of the Hagedorn mass spectrum $\rho(m)$ are well constrained by LQCD data in different sectors of strange quantum number, using the same limiting temperature $T_H\simeq 180 \, {\rm MeV} $. 
	
	The mass spectra for strange baryons inferred from the existing LQCD data are shown to be consistent with the trend of the unconfirmed resonances in the PDG. This is not the case for the strange-mesonic sector, where the corresponding $\rho(m)$ exceeds the current data of the PDG, even after the unconfirmed states are included. This may point to the existence of some uncharted strange mesons in the intermediate mass range. Clearly, new data and further lattice studies are needed to clarify these issues. Moreover, such missing resonances could be important for modeling particle production yields in heavy ion collisions.
	
	It would be interesting to assess the effects of resonance width on the Hagedorn spectrum. Recent studies suggest that the implementation of low-lying broad resonances in thermal models must be handled with care~\cite{Friman:2015zua, Broniowski}. The impact on the global spectrum and consequently the thermodynamics is currently under investigation. \\ \\ 
	
	\acknowledgements
	
	We acknowledge fruitful discussions with Bengt Friman. P.~M.~L. and M.~M. are grateful to E. Ruiz Arriola, W. Broniowski and M. Panero for their helpful comments and to M.~A.~R.~Kaltenborn for the careful reading of the manuscript. K.~R. also acknowledges fruitful discussion with A. Andronic, S. Bass, P. Braun-Munzinger, F. Karsch, M.  Nahrgang, J. Rafelski, H. Satz and J. Stachel and partial  support  of the  U.S. Department  of  Energy  under  Grant  No.  DE-FG02-05ER41367. C.~S. acknowledges partial support of the Hessian LOEWE initiative through the Helmholtz International Center for FAIR (HIC for FAIR). This work was partly supported by the Polish National Science Center (NCN), under Maestro Grant DEC-2013/10/A/ST2/00106.


\begin{thebibliography}{}
		
		\bibitem{BraunMunzinger:2003zd}
		P.~Braun-Munzinger, K.~Redlich and J.~Stachel,
		in {\it Quark-Gluon Plasma 3}, edited by R.~C.~Hwa {\it et al.} (World Scientific, Singapore, 2004), pp. 491-599
		\bibitem{tawfik1}
		F.~Karsch, K.~Redlich and A.~Tawfik,
		Eur.\ Phys.\ J.\ C {\bf 29}, 549 (2003).
		\bibitem{tawfik2}
		F.~Karsch, K.~Redlich and A.~Tawfik,
		Phys.\ Lett.\ B {\bf 571}, 67 (2003).
		\bibitem{reseqsf}
		F.~Karsch,
		Acta Phys.\ Polon.\ Supp.\  {\bf 7}, no. 1, 117 (2014).
		\bibitem{andronic}
		A.~Andronic, P.~Braun-Munzinger, J.~Stachel and M.~Winn,
		Phys.\ Lett.\ B {\bf 718}, 80 (2012).
		\bibitem{kapusta}
		M.~Albright, J.~Kapusta and C.~Young,
		Phys.\ Rev.\ C {\bf 90}, 024915 (2014).
		\bibitem{kapusta1}
		M.~Albright, J.~Kapusta and C.~Young,
		Phys.\ Rev. \ C {\bf 92}, 044904 (2015)
		\bibitem{gorenstein}
		V.~Vovchenko, D.~V.~Anchishkin and M.~I.~Gorenstein,
		Phys.\ Rev.\ C {\bf 91}, no. 2, 024905 (2015).
		\bibitem{Borsanyi:2011sw}
		S.~Borsanyi, Z.~Fodor, S.~D.~Katz, S.~Krieg, C.~Ratti and K.~Szabo,
		JHEP {\bf 1201}, 138 (2012).
		\bibitem{Bazavov:2012jq}
		A.~Bazavov {\it et al.}  [HotQCD Collaboration],
		Phys.\ Rev.\ D {\bf 86}, 034509 (2012).
		\bibitem{missing}
		A.~Bazavov, H.-T.~Ding, P.~Hegde, O.~Kaczmarek, F.~Karsch, E.~Laermann, Y.~Maezawa and S.~Mukherjee {\it et al.},
		Phys.\ Rev.\ Lett.\  {\bf 113}, no. 7, 072001 (2014).
		\bibitem{Borsanyi:2013bia}
		S.~Borsanyi, Z.~Fodor, C.~Hoelbling, S.~D.~Katz, S.~Krieg and K.~K.~Szabo,
		Phys.\ Lett.\ B {\bf 730}, 99 (2014).
		\bibitem{Majumder:2010ik}
		A.~Majumder and B.~Muller,
		Phys.\ Rev.\ Lett.\  {\bf 105}, 252002 (2010).
		\bibitem{Hagedorn:1965st}
		R.~Hagedorn,
		Nuovo Cim.\ Suppl.\  {\bf 3}, 147 (1965).
		\bibitem{Hagedorn:1971mc}
		R.~Hagedorn,
		CERN yellow report 71-12, (1971).
		\bibitem{raf1}
		J.~Letessier and J.~Rafelski,
		{\it Hadrons and Quark–Gluon Plasma},
		Cambridge Monographs on Particle Physics, Nuclear Physics and Cosmology, Vol. {\bf 18} (Cambridge University Press, Cambridge, 2005).
		\bibitem{Frautschi:1971ij}
		S.~C.~Frautschi,
		Phys.\ Rev.\ D {\bf 3}, 2821 (1971).
		\bibitem{Huang:1970iq}
		K.~Huang and S.~Weinberg,
		Phys.\ Rev.\ Lett.\  {\bf 25}, 895 (1970).
		\bibitem{Cudell:1992bi}
		J.~R.~Cudell and K.~R.~Dienes,
		Phys.\ Rev.\ Lett.\  {\bf 69}, 1324 (1992).
		\bibitem{Johnson:1975sg}
		K.~Johnson and C.~B.~Thorn,
		Phys.\ Rev.\ D {\bf 13}, 1934 (1976).
		\bibitem{Broniowski:2000bj}
		W.~Broniowski and W.~Florkowski,
		Phys.\ Lett.\ B {\bf 490}, 223 (2000).
		\bibitem{Broniowski:2004yh}
		W.~Broniowski, W.~Florkowski and L.~Y.~Glozman,
		Phys.\ Rev.\ D {\bf 70}, 117503 (2004).
		\bibitem{pdg}
		K.~A.~Olive {\it et al.}  [Particle Data Group Collaboration],
		Chin.\ Phys.\ C {\bf 38}, 090001 (2014).
		\bibitem{tcb}
		A. Bazavov {\it et al.} (HotQCD Collaboration),
		Phys. Rev. D {\bf 85}, 054503 (2012).
		\bibitem{tcw}
		Y. Aoki {\it et al.}, JHEP {\bf 0906}, 088  (2009).
		\bibitem{tc}
		T.~Bhattacharya, M.~I.~Buchoff, N.~H.~Christ, H.-T.~Ding, R.~Gupta, C.~Jung, F.~Karsch and Z.~Lin {\it et al.},
		Phys.\ Rev.\ Lett.\  {\bf 113}, no. 8, 082001 (2014).
		\bibitem{spectrum}
		Z.~Fodor and C.~Hoelbling,
		Rev.\ Mod.\ Phys.\  {\bf 84}, 449 (2012).
		
		\bibitem{Meyer:2009tq} 
		H.~B.~Meyer,
		Phys.\ Rev.\ D {\bf 80}, 051502 (2009).
		
		\bibitem{Caselle:2015tza} 
		M.~Caselle, A.~Nada and M.~Panero,
		JHEP {\bf 07}, 143 (2015).
		
		\bibitem{Arriola:2014bfa} 
		E.~Ruiz Arriola, L.~L.~Salcedo and E.~Megias,
		Acta Phys.\ Polon.\ B {\bf 45}, no. 12, 2407 (2014). 
		
		\bibitem{Arriola:2015gra} 
		E.~R.~Arriola, L.~L.~Salcedo and E.~Megias,
		Acta Phys.\ Polon.\ Supp.\  {\bf 8}, no. 2, 439 (2015).
		
		
		\bibitem{eqsf1}
		A.~Bazavov {\it et al.}  [HotQCD Collaboration],
		Phys.\ Rev.\ D {\bf 90}, no. 9, 094503 (2014).
		\bibitem{eqsf2}
		M.~Cheng, S.~Ejiri, P.~Hegde, F.~Karsch, O.~Kaczmarek, E.~Laermann, R.~D.~Mawhinney and C.~Miao {\it et al.},
		Phys.\ Rev.\ D {\bf 81}, 054504 (2010).
		\bibitem{ratti}
		C.~Ratti {\it et al.}  [Wuppertal-Budapest Collaboration],
		Nucl.\ Phys.\ A {\bf 855}, 253 (2011).
		\bibitem{ejiri1}
		S.~Ejiri, C.~R.~Allton, M.~Doring, S.~J.~Hands, O.~Kaczmarek, F.~Karsch, E.~Laermann and K.~Redlich,
		Nucl.\ Phys.\ A {\bf 774}, 837 (2006).
		\bibitem{ejiri2}
		S.~Ejiri, F.~Karsch and K.~Redlich,
		Phys.\ Lett.\ B {\bf 633}, 275 (2006).
		\bibitem{raf2}
		J.~Rafelski, J.~Letessier and A.~Tounsi,
		in {\it Hot Hadronic Matter: Theory and Experiment}, edited by J.~Letessier, H.~H.~Gutbrod, and J.~Rafelski, NATO ASI, Series B: Physics, Vol. 346 (Springer, Berlin, 1995), pp. 479-492.
		\bibitem{satz}
		K.~Redlich and H.~Satz,
		in {\it Melting Hadrons, Boiling Quarks - From Hagedorn Temperature to Ultra-Relativistic Heavy-Ion Collisions at CERN}, edited by J.~Rafelski, (Springer International Publishing, 2016), pp. 49-68. 
		\bibitem{Friman:2015zua} 
  		B.~Friman, P.~M.~Lo, M.~Marczenko, K.~Redlich and C.~Sasaki,
  		Phys.\ Rev.\ D {\bf 92}, 074003 (2015). 
		\bibitem{Broniowski} 
		W.~Broniowski, F.~Giacosa and V.~Begun,
		Phys.\ Rev.\ C\ {\bf 92}, 034905 (2015)
		
		
	\end{thebibliography}
\end{document}